\documentclass{mbe}
\usepackage{amsmath}
  \usepackage{paralist}
  \usepackage{graphics} 
  \usepackage{epsfig} 
\usepackage{graphicx}  \usepackage{epstopdf}
\usepackage{multirow}
\usepackage{xcolor}
\usepackage{booktabs, tabularx, colortbl}
 \usepackage[colorlinks=true]{hyperref}
\hypersetup{urlcolor=blue, citecolor=red}

\def\currentvolume{XX}
 \def\currentissue{XX}
  \def\currentyear{2016}
   \def\currentmonth{}
    \def\ppages{X--XX}
     \def\DOI{10.3934/mbe.201X.XX.xx}

\theoremstyle{definition}

\newcommand{\beq}{\begin{equation}}
\newcommand{\eeq}{\end{equation}}

\newcolumntype{a}{>{\columncolor{white}}c}
\newcolumntype{b}{>{\columncolor{gray!30}}c}
\newcommand{\R}{\mathcal{R}_0}

\usepackage[normalem]{ulem}

\usepackage[normalem]{ulem}

\usepackage[normalem]{ulem}

\title[Analysis of a Small Ebola Outbreak] 
      {Network-Based  Analysis of a Small Ebola Outbreak}

\author[Burch, Jacobsen, Tien, and Rempala]{}


 \email{burch.126@osu.edu}
 \email{jacobsen.50@osu.edu}
 \email{jtien@math.ohio-state.edu}
 \email{rempala.3@osu.edu}

\thanks{This research has been supported in part by the Mathematical Biosciences Institute and the National Science Foundation under grants RAPID DMS-1513489 and DMS-0931642.}

\begin{document}
\maketitle

\centerline{\scshape Mark G. Burch}
\medskip
{\footnotesize
 \centerline{College of Public Health}
   \centerline{The Ohio State University}
   \centerline{Columbus, OH 43210, USA}
} 

\medskip

\centerline{\scshape Karly A. Jacobsen}
\medskip
{\footnotesize
 \centerline{Mathematical Biosciences Institute}
   \centerline{The Ohio State University}
   \centerline{Columbus, OH 43210, USA}}

   \medskip

\centerline{\scshape Joseph H. Tien}
\medskip
{\footnotesize
 \centerline{Department of Mathematics and Mathematical Biosciences Institute}
   \centerline{The Ohio State University}
   \centerline{Columbus, OH 43210, USA}}

      \medskip

\centerline{\scshape Grzegorz A. Rempa{\l}a}
\medskip
{\footnotesize
 \centerline{Mathematical Biosciences Institute and College of Public Health}
   \centerline{The Ohio State University}
   \centerline{Columbus, OH 43210, USA}}

\bigskip


\begin{abstract}
We present a method for estimating epidemic parameters in network-based stochastic epidemic models when the total number of infections is assumed to be small. We illustrate the method by reanalyzing the   data from the 2014 Democratic Republic of the Congo (DRC) Ebola outbreak described in Maganga et al. (2014). \end{abstract}

\section{Introduction}

The best known models for the spread of infectious disease in human populations are based on the classical SIR model of Kermack and McKendrick \cite{Kermack1927}. The same system of ordinary differential equations (ODEs) may be derived as the large population limit of a density-dependent Markov jump process using the methods of Kurtz \cite{Kurtz1971}. This stochastic formulation brings a number of mathematically attractive properties such as explicit likelihood formulas and ease of simulation. Nevertheless, a drawback of the Kermack and McKendrick-type models is that they can be unrealistic in describing the interactions of infectives and susceptibles as they are based on assumptions of homogeneous mixing \cite{Keeling2005}. 

In recent years there has been considerable interest in developing alternatives to the classical SIR, for instance, via network-based epidemic models, as reviewed by Pellis et al. \cite{Pellis2015} and House and Keeling \cite{House2011}.  Pair approximation models have been formulated to account for spatial correlations while maintaining mathematical tractability \cite{Rand1999,Keeling1999,Keeling1999b}. Much of the work on these models has regarded the dynamics of the deterministic system obtained in the large population limit, such as the work of Miller and Volz on edge-based models \cite{Volz2008,Miller2011,Miller2013} and that of Altmann \cite{Altmann1998} on a process with dynamic partnerships. 

As in the work of Miller and Volz \cite{Volz2008,Miller2011,Miller2013}, the configuration model (CM) random graph is often chosen to dictate the network structure in epidemic models \cite{Barbour2013, Meloni2011,Boguna2013}.  CM networks can be viewed as a generalization of the Erd{\"o}s-R{\'e}nyi random graph, i.e. the $G(n,p)$ model where $n$ is the size of the graph and edges are drawn between each pair of nodes independently, with probability $p$.  In the limit of a large graph this results in a Poisson degree distribution; however, in a general CM graph model framework Poisson may be replaced with any appropriate degree distribution (see, e.g., \cite{Volz2008}).  Miller and Volz heuristically formulated the limiting system of ODEs that govern the dynamics of the SIR epidemic on a CM graph.  Decreusefond et al. \cite{Decreusefond2012} and Janson et al. \cite{Janson2014} have recently given formal proofs for the correctness of the Miller-Volz equations as the law of large numbers (LLN) for the stochastic system under relatively mild regularity conditions on the degree distribution and on the epidemic initial condition.  In addition, Volz and Miller have conjectured  the limiting equations for generalizations which include dynamic graphs (edge formation and breakage), heterogeneous susceptibility, and multiple types of transmission \cite{Miller2013}. To date, the convergence for the stochastic systems in these cases has not been formally verified but numerical studies done by the authors suggest that they are correct. 

As in the case of classical SIR \cite{allen2008introduction} the early behavior of an epidemic on a CM graph can be approximated by a suitable branching process \cite{Barbour2013,Janson2014,Bohman2012}. Such approximation  allows one in turn to use the early epidemic data to statistically ascertain the probability of a major outbreak (large number of infections among the network nodes)  as well as to estimate the rate of infection spread and changes in the contact network as described below.   
By and large, there have been limited studies of statistical estimation for network-based models \cite{Welch2011}. In one of the early papers on the topic, O'Neill and Britton  study Bayesian estimation procedures for the $G(n,p)$ model when the network is assumed to be of reasonable enough size that the network structure may be included as missing data and imputed via a Monte Carlo scheme \cite{Britton2002}. More recently, Groendyke has extended this approach to non-Markovian dynamics by allowing both the infection time and the recovery time to  follow  an arbitrary gamma distribution \cite{Groendyke2011,Groendyke2012}. However, these  methods are generally  tailored to networks of small size, where posterior sampling is feasible, and  may encounter various difficulties in  large networks when  the imputation  becomes computationally expensive and often impractical.  The framework presented here, on the other hand, assumes the proportion of infectives is small relative to the population size.  This allows us to avoid explicit imputation of the network and makes the numerical complexity of the analysis comparable to that of a small homogenous  SIR epidemic  \cite{Choi:2012aa,Schwartz:2015aa}.

The main contribution of the current paper is to present a statistical inference method for analyzing the early stages of an epidemic, or a small outbreak, evolving according to SIR type dynamics on a random graph.  A novel aspect of our method is that we assume the random graph structure evolves in response to the epidemic progression.  This allows us to account for changing contact patterns in response to infection, for example due to population behavioral changes or health interventions (e.g. quarantine).

The development of our methods is motivated by the devastating $2013-2015$ outbreak of Ebola virus in West Africa.  While the West Africa outbreak received considerable attention in fall 2014 due to the dramatic rise in the number of cases, there was an independent Ebola outbreak in the Democratic Republic of the Congo (DRC) occurring at the same time. This much smaller outbreak began July 26, 2014 in {\'E}quateur Province and lasted until November 2014.  The index case was a woman living in Inkanamongo village that presumably became infected by consuming bushmeat of an infected animal. Several healthcare workers involved in a postmortem cesarean section on the women subsequently became ill and generated further chains of transmission \cite{Maganga2014}.


\section{DRC dataset} \label{data}

\begin{figure}[t]
	\begin{center}
		\includegraphics[scale=.6]{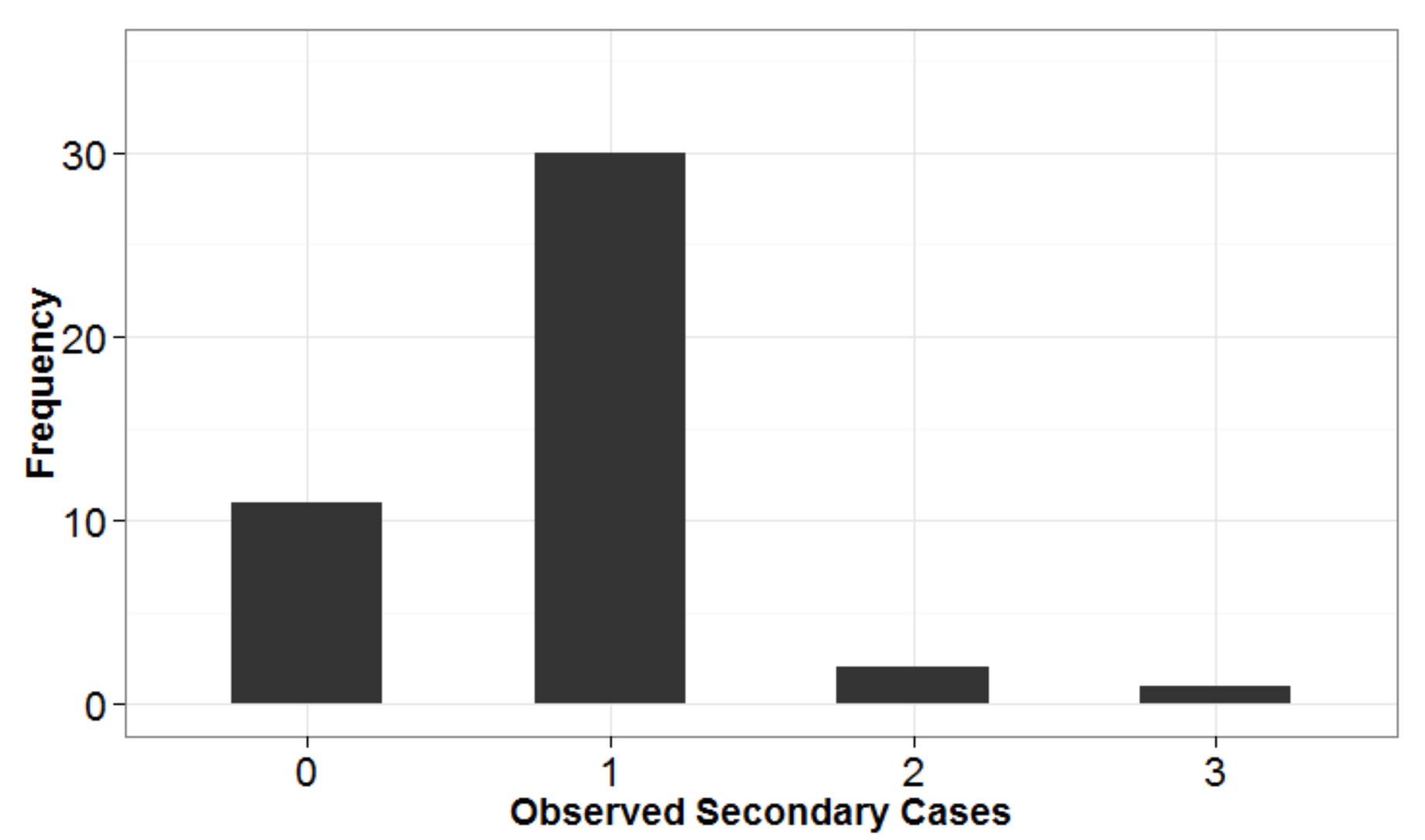}
		\caption{The empirical secondary case distribution in  the DRC outbreak dataset (neglecting the index case), as given by Maganga et al. \cite{Maganga2014}.}
		\label{secondcases}
	\end{center}
\end{figure}

There were a total of $69$ confirmed infections in the 2014 DRC outbreak.  The time series of cumulative case counts was reported by Maganga et al. \cite{Maganga2014}.  However, the analysis method presented here does not require temporal data and instead utilizes the distribution of secondary cases, also given in \cite{Maganga2014}.  The index case is believed to have caused $21$ secondary cases, presumably due to her funeral acting as a super-spreading event \cite{Maganga2014}.  Hence, we assume this data point to be an outlier and exclude it from our final analysis, as Maganga et al. also did in their estimation of $\R$.  The secondary case distribution for other named contacts is shown in Figure \ref{secondcases}.  One patient caused three subsequent cases, two patients caused two additional cases, $30$ caused one additional case and $11$ patients caused zero additional cases. These may be viewed as observations from the post-index offspring distribution of the branching process approximation, i.e. the number of infections caused by an individual who is himself not the index case. 

Although, as already mentioned,   the  temporal data measurements are not explicitly required for parameter estimation in our approach,   the available temporal  information can  be directly incorporated into the likelihood function,  or used to inform the parameterization of the prior distributions, which is  what was   done for the current DRC dataset. Details are given  below.

\section{Epidemic model} \label{model}
Consider a graph $G=\langle V,E \rangle$ where the vertex set, $V$, corresponds to individuals in a population of size $n$ and the edge set, $E$, corresponds to potentially infectious contacts occurring between such individuals.   The graph structure is given as a realization of a configuration model random graph with a prescribed degree distribution, $D$, where the probability that a node has degree $k$ is denoted by $q_k  \equiv P(D=k)$.  Each node $i \in V$ is assigned $d_i$ half-edges that are drawn at random from $D$. Then, the pool of half-edges is paired off uniformly to form the final network. This link formation model does not exclude the possibility of  self-loops or multiple edges but  this has a negligible effect in the large graph limit (see, e.g., discussion in \cite{Janson2014}). 

The general disease framework adopted is the standard compartmental SIR model where, for every time $t>0$, each node $i \in V$ is classified as susceptible (S), infectious (I), or recovered (R) from the infection. Given some number of initially infectious nodes, a node $i$ becomes infective via transmission along an edge from one of his infectious neighbors. In the Markovian case, we assume that $i$ remains infectious for an exponentially distributed amount of time with rate parameter $\gamma$, which we refer to as the recovery rate. More generally, a  non-exponential (e.g., gamma) recovery rate  leads to the so-called  semi-Markov SIR model \cite{janssen2013semi}.  While infectious, the infective $i$ attempts to transmit the infection to all of his susceptible neighbors according to an exponential distribution with rate $\beta$. In the case that the realization of the infection ``clock" for a particular neighbor ``rings" before the recovery ``clock", then this neighbor becomes infected. The epidemic ends when there are no more infectives. 

In addition to these standard SIR dynamics, we allow the network structure to change due to infection status.  Specifically, we assume that infectious individuals drop each of their contacts according to an exponential distribution with rate $\delta$.  The dropped contacts, which could account for behavioral changes due to disease such as isolation or decreased mobility, cannot be reformed. 

\section{Statistical inference} \label{inference}
\subsection{Index case offspring distribution}
For an index case $i$ with degree $d_i$, at most $d_i$ secondary infections can be produced. Conditional on the recovery time of that index case, $t_i$, the probability that infection has passed to any particular neighbor is given by 
\begin{equation}
p_{t_i} \equiv p(t_i;\beta,\delta) = \frac{\beta }{\beta+\delta}(1-e^{-(\beta + \delta) t_i}) \label{eq:pt}
\end{equation}
where the first term in the product on the right-hand side is the probability that an edge transmitted infection prior to being dropped, while the second term represents the probability that either an infection or a drop occurred before recovery.  Therefore, the total number of secondary infections caused by node $i$, which we denote by $X_i$, conditional on the time of recovery $t_i$, is given by 
\begin{equation*}
P(X_i = x_i | t_i, d_i) = \dbinom{d_i}{x_i}p_{t_i}^{x_i} (1-p_{t_i})^{d_i-x_i}.
\end{equation*}
If the recovery time is not known but assumed to follow a distribution $f(t_i) \equiv P(T_i= t_i)$, then we may analogously define $\pi_{d_i}(x_i)$, the conditional probability of $x_i$ offspring given degree $d_i$.  The law of total probability implies

\begin{equation*}
\pi_{d_i}(x_i) = \int_{0}^{\infty} \dbinom{d_i}{x_i}p(t_i)^{x_i} (1-p(t_i))^{d_i-x_i} f(t_i)dt_i.
\end{equation*}
We will assume identical recovery distributions for all individuals and, thus, the offspring distribution will be the same for all index cases. 

The final form for the offspring distribution of an index case may be found by supposing that the degree, $d_i$, of the index case is unknown. Let $q_{d_i} \equiv P(D_i = d_i)$ denote the probability that the index case has degree $d_i$. Therefore, the law of total probability gives
\begin{equation}
P(X_i = x_i) = \sum_{d_i \geq x_i} q_{d_i} \pi_{d_i}(x_i). \label{offspring1}
\end{equation}
That is, we sum over all possible degrees that could yield at least $x_i$ secondary infections and weight them according to the degree distribution.

\subsection{Post-index case offspring distribution}

We now consider the offspring distribution for a post-index case.  By definition, such an individual acquired infection from another individual in the network and, thus, has at least one neighbor.  Therefore, some adjustments are needed to account for the fact that post-index cases have a degree distribution which differs from $D$. Let $q'_k$ denote the probability that a given neighbor in the CM network has degree $k$. Then it is known \cite{Newman2010} that
\begin{equation*}
q'_k = \frac{kq_k}{\mu}.
\end{equation*}
Since at least one of the neighbors of a post-index case has already been infected, he may pass the infection to at most $k-1$ of his neighbors. Let $X'_i$ denote the offspring distribution for a post-index infection $i$.  Similarly to Eq. $(\ref{offspring1})$ we derive
\begin{align}
P(X'_i = x'_i) = \sum_{k > x'_i}^{\infty} q'_k \pi_{k-1}(x'_i). \label{offspring2}
\end{align}

For a fixed set of parameters the basic reproductive number, $\R$, can be calculated as $E(X'_v)$, i.e. the average number of secondary infections caused by a post-index case.  That is, $\R$ is given by  
 \begin{equation}
\R = \sum_{x'_i=0}^\infty x'_i \sum_{k > x'_i}^{\infty} q'_k \pi_{k-1}(x'_i). \label{eq:R0}
\end{equation}

\subsection{Example} \label{sec:ex}

To illustrate, we assume that the degree distribution is Poisson with mean parameter $\lambda$ and the recovery distribution is exponential with rate parameter $\gamma$.  This implies
\begin{equation*}
f(t_i) = \gamma e^{-\gamma t_i}
\end{equation*}
and
\begin{equation*}
q_{d_i} = \frac{\lambda^{d_i} e^{-\lambda}}{d_i!}.
\end{equation*}
Therefore, by Eq. $(\ref{offspring1})$, the offspring distribution for an index case is given by
\begin{equation*}
P(X_i = x_i|\lambda, \beta, \gamma, \delta) = \sum_{k \geq x_i}^{\infty} \frac{\lambda^{k} e^{-\lambda}}{k!} \int_{0}^{\infty} \dbinom{k}{x_i}p(t; \beta, \delta)^{x_i} (1-p(t; \beta, \delta))^{k-x_i} \gamma e^{-\gamma t} dt,
\label{offspringexp1}
\end{equation*}
and, by Eq. $(\ref{offspring2})$, the post-index case offspring distribution is given by
\begin{align*}
P(X'_i = x'_i &|\lambda, \beta, \gamma, \delta) \nonumber \\ 
&= \sum_{k > x'_i}^{\infty} \frac{\lambda^{k-1} e^{-\lambda}}{(k-1)!} \int_{0}^{\infty} \dbinom{k-1}{x'_i}p(t; \beta, \delta)^{x'_i} (1-p(t; \beta, \delta))^{k-1-x'_i} \gamma e^{-\gamma t} dt. 
\end{align*}

This expression has no simple analytical form but it is not hard to approximate the integral numerically since it can be written as an expectation against the recovery distribution. Therefore, a simple Monte Carlo sample from the desired recovery distribution allows for efficient computation of this term.

Using Eq. $(\ref{eq:R0})$, we can calculate the basic reproductive number in this Markovian case.  For an arbitrary degree distribution, $\R$ is given by
\begin{equation}
\R = \dfrac{\beta}{\beta+\gamma+\delta}\sum_{k=0}^\infty \dfrac{(k-1)kq_k}{\mu}, \label{R0eqn}
\end{equation}
which only differs in the inclusion of $\delta$ from the corresponding formula on a static CM graph \cite{Miller2011,Janson2014}. Here $\mu=E(D)<\infty$ by assumption.  The summation in Eq. $(\ref{R0eqn})$ represents the expected excess degree, i.e. the degree of a node which is necessarily a neighbor of a node, not counting the known edge.  In the particular case here of a Poisson degree distribution, $\R$ is found to be (cf., e.g.,  \cite{Andersson2000} chapter 6)
\begin{align}
\R &= \frac{\beta}{\beta +\gamma+\delta} \sum_{k=0}^\infty \frac{\lambda^{k-1} e^{-\lambda}}{k!} k(k-1) = \frac{\beta\,\lambda}{\beta +\gamma+ \delta}. \label{eq:R0ex}
\end{align}

\subsection{Likelihood and estimation}
In practice, outbreak data may not arise solely from a single index case and often $m$ separate chains of infection are tracked. Suppose the data $\{x_1, ... ,x_m\}$ corresponds to the number of secondary infections for each of $m$ independent index cases and the data $\{x'_1,  ..., x'_{m'}\}$ corresponds to secondary infections caused by each of the $m'$ post-index cases.  

Let $\Theta = (\beta,\gamma,\delta,\lambda)$ denote the vector of parameters where $\gamma$ and $\lambda$ represent the parameters of the recovery time and degree distributions, respectively.  The offspring distributions given in Eqs. $(\ref{offspring1})$ and $(\ref{offspring2})$ allow for explicit formulation of the likelihood for $\Theta$ which is given by 
\begin{align}\label{eq:lkhd}
L(\Theta &| x_1, ... , x_m, x'_1, ..., x'_{m'}) = \prod_{j=1}^{m} \sum_{k \geq x_j}^{\infty} q_k \pi_k(x_j) \times \prod_{c=1}^{m'} \sum_{l > x'_c}^{\infty} q'_l \pi_{l-1}(x'_c).
\end{align}


With the specification of the likelihood, maximum likelihood estimators (MLEs) for the rate parameters can be found by numerical optimization.  Given the MLE $\hat{\theta} = (\hat{\beta},\hat{\gamma},\hat{\delta},\hat{\lambda})$, the corresponding estimator for the basic reproductive number can be calculated by application of the continuous mapping theorem to the expression for $\R$ given in Eq. $(\ref{eq:R0})$. For example, under the assumptions of our example in Section \ref{sec:ex}, the estimator following from Eq. $(\ref{eq:R0ex})$ would be
\begin{equation}
\hat{\R} = \dfrac{\hat{\beta}\,\hat{\lambda}}{\hat{\beta}+\hat{\gamma}+\hat{\delta}}.
\end{equation}

Denote the vector of current parameters as $\Theta = (\beta, \gamma, \delta, \lambda)$.  If we denote the parameter prior distribution  $\phi(\Theta)$ then the likelihood function above may be also used to  compute the Metropolis-Hastings acceptance probability in the  Markov chain Monte Carlo (MCMC) sampler.    
Let $\boldsymbol{x}$ denote the vector of data counts and $\tau$ be the transition kernel. The MCMC algorithm for obtaining the posterior distribution of $\Theta$ is then as follows
\begin{enumerate} \label{MCMC}
\item Initiate $\Theta_{curr} = \Theta_0$.
\item Obtain proposal $\Theta_{prop}$ from $\tau(\Theta|\Theta_{cur})$.
\item Accept (or not) this proposal with Metropolis-Hastings probability given by
\begin{equation}
\rho(\boldsymbol{x}, \Theta_{cur}, \Theta_{prop})= min\left(1, \frac{L(\Theta_{prop}|\boldsymbol{x}) \phi(\Theta_{prop}) \tau(\Theta_{prop}|\Theta_{cur})}{L(\Theta_{cur}|\boldsymbol{x}) \phi(\Theta_{cur}) \tau(\Theta_{cur}|\Theta_{prop})}\right).  
\end{equation}
\item Return to 2.
\item Repeat until convergence.
\end{enumerate}

In this way, after sampler convergence, we obtain an approximate sample from the  posterior distribution of $\Theta$ and  may subsequently compute its  approximate  $1-\alpha$ credibility  region,  given the observed   data $\boldsymbol{x}$.  In particular, the credibility  interval of the parameter $\R$ given by Eq. \eqref{eq:R0} may be    determined.  

\subsection{Generalizations} \label{general}
Note that the likelihood formula \eqref{eq:lkhd} is valid in a non-Markovian setting such as an arbitrary recovery time distribution and could further be extended to the scenario where the transmission rate varies with time since infection.  Note that in the latter case the formula for $\R$ would differ from Eq. \eqref{eq:R0} due to a form for $p_{t_i}$ that differs from Eq. \eqref{eq:pt} but would remain calculable as $E(X'_v)$. 

If additional data were available, such as the recovery time or number of contacts of each individual, it could be explicitly incorporated into the likelihood function \eqref{eq:lkhd} through the joint distribution of recovery times and degrees.

\section{Analysis of the  DRC dataset} \label{results}

To illustrate our method we perform the Bayesian posterior estimation of the model parameters for the 2014 Ebola outbreak in the DRC based on the data described in Section~\ref{data}.  The specific model considered is as given in Section~\ref{model}, where transmission occurs according to an exponential distribution with rate parameter $\beta$, and the degree distribution is Poisson with parameter $\lambda$.  Recovery time is assumed to follow a gamma distribution $\Gamma(\alpha, \beta)$.  We note that, while incubation periods for Ebola range from two to 21 days \cite{Chowell2014b}, our method does not require consideration of latent exposure since it does not depend on infection timing.

\begin{figure}[t]
	\begin{center}
		\includegraphics[scale=.55]{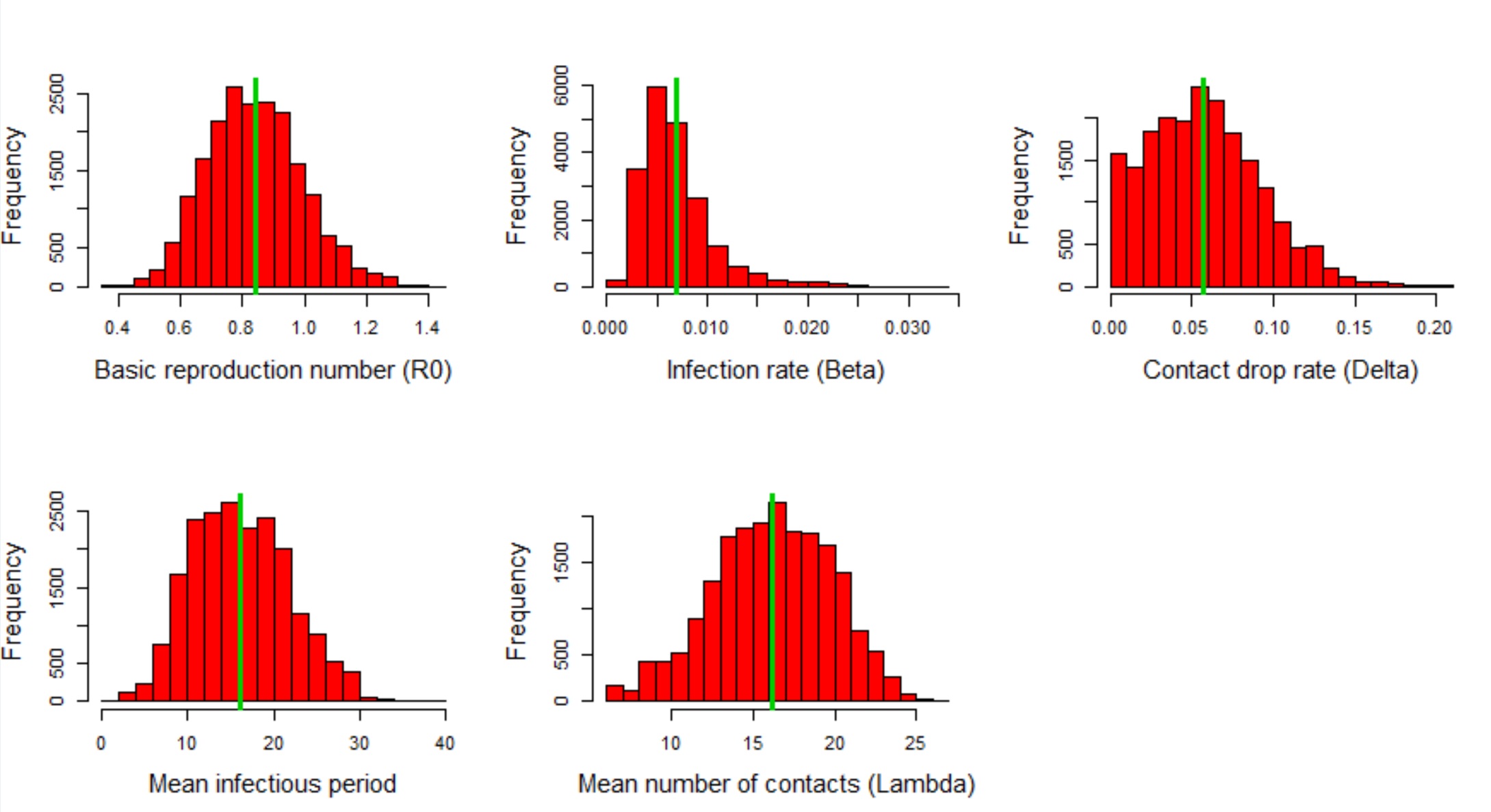}
		\caption{Estimated posterior densities for the parameters of interest for the non-Markovian MCMC sampler.  Green line denotes posterior mean.}
		\label{posteriors}
	\end{center}
\end{figure}

We perform estimation via the MCMC scheme given in Section \ref{MCMC}. Prior distributions were set to be minimally informative Gaussian distributions and hyper-parameters were selected based on previous estimates \cite{Maganga2014, Team2014}. The prior distribution for $\lambda$ was taken to be $N(16,14)$ and for $\delta$ was taken to be $N(.01, .05)$. To improve mixing of the MCMC scheme, $\beta$ was estimated on log-scale under an assumed $N(\log(.02),4)$ prior distribution. Lastly, the gamma distribution for the infectious period was chosen to have prior mode (i.e., $(\alpha-1)/\beta$) distributed as $N(11,6)$ and prior standard deviation ($\sqrt{\alpha}/\beta$) as $N(6,4)$. A Gaussian transition kernel $\tau$ was used. Central $95\%$ credibility intervals were calculated for the parameters of interest as well as for $\R$. Results are summarized in Figure \ref{posteriors} including histograms of posterior samples. 

 The   $95\%$ credibility interval for the  basic reproductive number $\R$ is found to be $(.589, 1.15)$ with the $\R$ posterior mean of $.842$. The latter value  numerically agrees closely with the moment-based estimate given  in  Maganga et al. \cite{Maganga2014}  and is also consistent with the relatively small total number of confirmed infections observed during the epidemic. We further  note that our interval estimate compares favorably  to the  $95\%$  $\R$ confidence interval of $(-.38, 2.06)$  reported by Maganga et al. \cite{Maganga2014} indicating  that for the DRC data  the fully parametric model  produces a more precise (shorter)  interval. The infection rate is found to have a posterior mean of $.0069$ and a corresponding 95\% credibility interval of $(.00232,.0167)$. The contact drop rate is found to have a posterior mean of $.0573$ and corresponding credible interval of $(.00340,.128)$. These quantities were not estimated directly by the  authors in   \cite{Maganga2014} so  comparison here is not possible. The estimated mean for the infectious period is found to be $16.18$ days with corresponding credibility interval of ($6.65$ days, $27.9$ days).  Maganga et al. did not explicitly estimate infectious period but instead gave an estimate for time from symptom onset to death with mean 11.3 days \cite{Maganga2014}.  Based on a historic 1995 outbreak of Ebola in the DRC, Legrand et al. estimated the time from onset to end of infectiousness for survivors to be 10 days and time from onset to death to be 9.6 days \cite{Legrand2007}.  While our estimate of infectious period is somewhat longer than these, we note that our interpretation of the quantity is the total length of time that an individual could cause infection.  This could include several days after death in which the body is being prepared for and undergoing funeral rituals.  Lastly, the mean number of contacts is estimated to be $16.2$ with corresponding credibility interval of $(8.59, 22.8)$, which is again in close pointwise agreement with the estimate inferred from the number of contacts traced by Maganga et al. \cite{Maganga2014}. Overall, the point estimate for $\R$ and infection and recovery rates are  seen to be consistent with a small outbreak behavior observed in the DRC dataset and to agree well with the numerical values reported earlier. However,  based on the same data,    our  parametric model is also seen to yield more precise interval estimates.
 
\begin{figure}[t]
 	\begin{center}
 		\includegraphics[scale=.75]{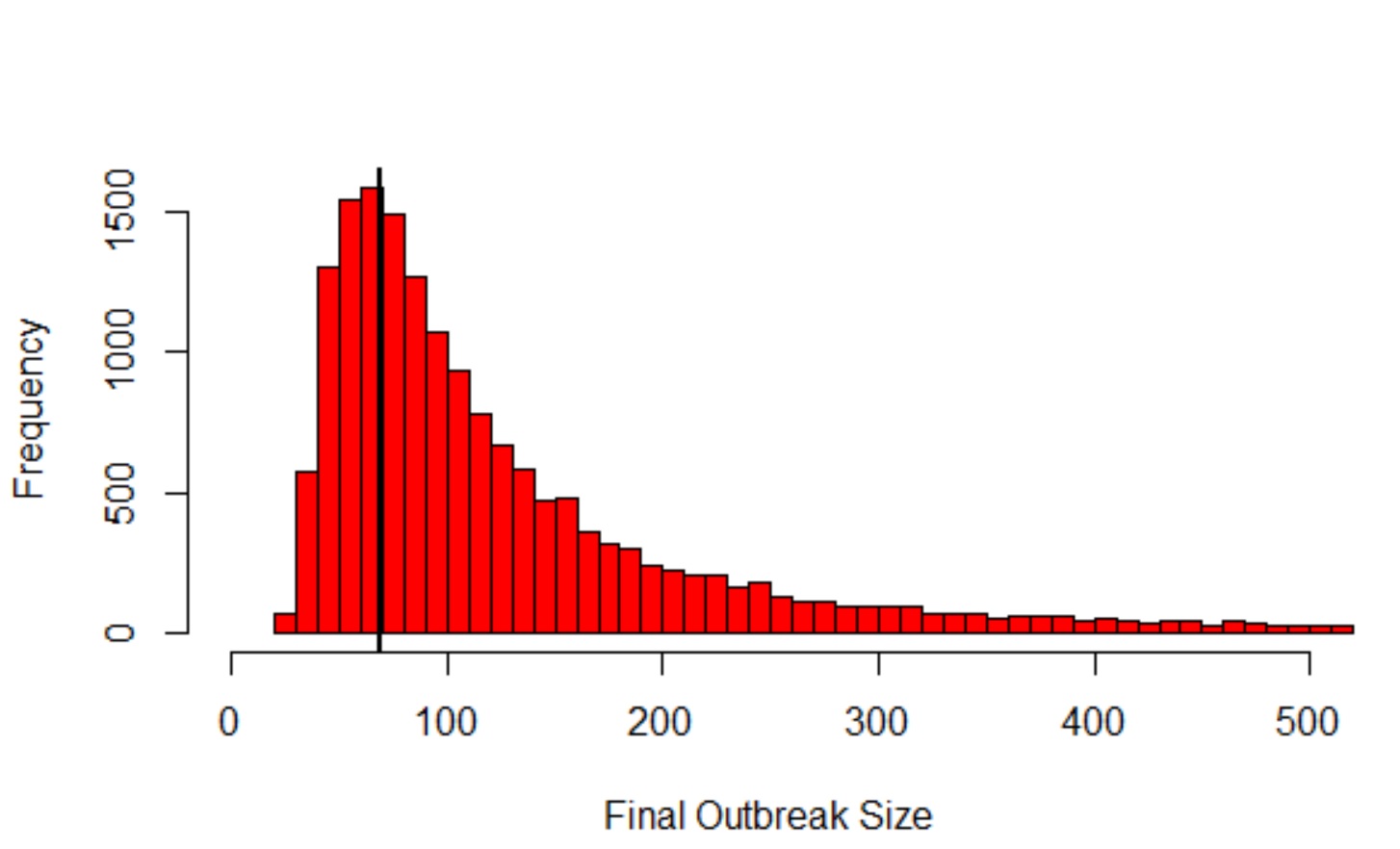}
 		\caption{Final outbreak size distribution based on 20,000 simulations of the branching processes from the posterior parameter distribution.  The actual outbreak size based on  the DRC dataset (black line) is  shown for comparison.}
 		\label{fig:finalsize}
 	\end{center}
 \end{figure}

The final outbreak size of simulated branching processes from the posterior parameter distribution is used as a model diagnostic for fit to the empirical data.  Conditioning on the number of infections caused by the index case as the number of independent branches, the posterior parameter samples and corresponding post-index case offspring distributions were used to simulate branching process realizations.  The final outbreak sizes were calculated and the distribution is presented in Figure \ref{fig:finalsize}.  Reasonable agreement is observed between this distribution and the empirical outbreak size of 69 cases \cite{Maganga2014}.

\section{Discussion} \label{discussion}

We presented here a Bayesian parameter estimation method for a class of stochastic epidemic models on configuration model graphs.  The method is based on applying the branching process approximation, and is applicable when the number of infections is small in relation to the size of the population under study.  In particular,  this includes the case when the total outbreak size is small or when  we are at the onset of a large outbreak.  The  method are flexible, for example allowing for arbitrary degree and recovery distributions, and in principle requires only a knowledge of the distribution of secondary cases, although   additional data can be incorporated into the inference procedure, as the likelihood  function under branching approximation remains straightforward to evaluate under a wide range of  data collection schemes.

We illustrated our approach   with the analysis of data from the 2014 DRC Ebola outbreak which was originally described and analyzed  in Maganga et al. \cite{Maganga2014}. Our method, under only weakly informative prior distributions, is seen to produce a considerably tighter credibility interval for $\R$ than  the moment-based confidence estimate reported in  \cite{Maganga2014}. This demonstrates the utility of the branching process approximation for small epidemics in obtaining   more precise  estimates of  $\R$, which is essential in assessing potential risk of a large outbreak and in determining the level of control efforts (e.g. vaccination or quarantine) needed to mitigate an outbreak.  The final size comparison indicates that the observed data is within the range of model predictions, although   the direct comparison  of  observed and model predicted offspring distributions indicates some disagreement in the observed frequency of zeros and ones.  The small sample size prohibits definitive conclusions but this may indicate the need to incorporate more complex network dynamics (e.g. distinguishing between multiple types of infectious contacts). 

As  the  statistical estimation methods appear  essential  to  inform public health interventions,  we hope 
that  our work  here  will help in establishing  a  broader  inference  framework  for  epidemic parameters,  based on  the type of data usually  collected in the course of an outbreak.   The Bayesian approach is particularly attractive in this context, as it naturally  incorporates any prior or historical information.  However, the current approach  only addresses the inference problem at the epidemic onset and, in particular, is not appropriate when the number of infected individuals comprises a significant portion of the population. We plan to address estimation for such large outbreaks,  possibly  also  incorporating  more complex network dynamics, in our future work.

\section*{Acknowledgments} We thank the Mathematical Biosciences Institute at OSU for its assistance in  providing us with space and the necessary  computational resources.


\providecommand{\href}[2]{#2}
\providecommand{\arxiv}[1]{\href{http://arxiv.org/abs/#1}{arXiv:#1}}
\providecommand{\url}[1]{\texttt{#1}}
\providecommand{\urlprefix}{URL }

\end{document}